# Color breaking in the quantum leaped stop decay

**Imre Czövek**

imre.czovek@gmail.com

**Abstract**. The superfield propagator contains a measurable quantum leap, which comes from the definition of SUSY.
In the sfermion -> Goldstino + fermion vertex change:

1. the spin of sparticle with discrete 1/2,
2. the Grassman superspace with the Goldstino shift operator.
3. the spacetime as the result of extra dimensional leap.

The leap nature of SUSY transformations appears in the squark decay, it is the analog definition of SUSY. The quantum leaped outgoing propagators are determined and break locally the energy and the charge. Like to the teleportation the entangled pairs are here the b quark and the Goldstino.
The dominant stop production is from gluons. The stop-antistop pair decay to quantum leaped b (c or t) quark, and the decay break the color.
I get for the (color breaking) quantum leap: 10^-18 m.
10^-11 m color breaking would be needed for a color breaking chain reaction.
The open question is: Are the colliders producing supersymmetry charge? Because some charges in QGP can make long color breaking and a chain reaction.

A long color broken QGP state in the re-Big Bang theory could explain the near infinite energy and the near infinite mass of the universe:
- at first was random color QGP in the flat space-time,
- at twice the color restoration in the curved space-time, which eats the Goldstinos,
- and finally the baryon genesis.

The re Big Bang make a supernova like collapse and a flat explosion of Universe.

This explanation of SUSY hides the Goldstone fermion in the extra dimensions, the Goldstino propagate only in superspace and it is a not observable dark matter.

## 1. Supersymmetry introduction

Bosonical superfield in SUSY:

$$\phi = A + \sqrt{2}\Theta\psi + \Theta\Theta F \tag{1.1}$$

The Q SUSY generator mixes the original state with its super partner state.

$$\psi \xrightarrow{\varepsilon Q} \psi + \bar{\varepsilon} i \sigma^{\alpha} \partial_{\alpha} A \tag{1.2}$$

The definition of two infinitesimal SUSY transformations is:

$$[\delta_1, \delta_2]\Psi = \left[\varepsilon_1 \overline{Q}, \overline{\varepsilon_2 Q}\right]_{-} \Psi = \bar{\varepsilon}_2 2i\sigma^{\mu}\varepsilon_1 \partial_{\mu}\Psi = a^{\mu}\partial_{\mu}\Psi = \delta_a \Psi \tag{1.3}$$

The space-time translation is the same for fermions and bosons. The 4D space shift with extra dimensional Grassman shifts in SUSY definition:

$$a^{\mu} = \bar{\varepsilon}_2 2i\sigma^{\mu}\varepsilon_1 \tag{1.4}$$

The space-time shift of particles was defined with extra dimensional shift-operators. The Goldstone fermion is a candidate for Grassman shifts operator in spontaneous supersymmetry breaking theorem.
If the Goldstino shifts the Grassman space in both Grassman dimensions, then this SUSY shift in eq. 1.3. is hermetian and causal.

$$\Theta_2 = \Theta_1 + \tilde{G}_{\Theta}, \quad \overline{\Theta}_2 = \overline{\Theta}_1 + \overline{\tilde{G}}_{\overline{\Theta}} \tag{1.5a}$$

$$\tilde{G} = \frac{1}{\sqrt{2}}(\tilde{G}_{\Theta} + \overline{\tilde{G}}_{\overline{\Theta}}) \tag{1.5b}$$

Goldstino field is a continuous parameter in Grassman space.

$$\tilde{G} = \sum_{i=1}^{\infty} \varepsilon_i \tag{1.6}$$

Eq 1.4. quantum shift was written in many books, but they assumed that it is infinitesimal small and so continuous. But the antikommutator in eq. 1.3. is not dissolvable, like to field. (The product of linear combinations is not equivalent with linear combinations of products.) The quantum leap is a discrete constant, isn't a continuous translation, and the antikommutator is not dissolvable to the sum of measurable infinitesimal parts.

$$[\tilde{G}\overline{Q}, \overline{\tilde{G}}Q]_{-} \neq \sum[\delta\varepsilon\overline{Q}, \delta\bar{\varepsilon}Q]_{-} \tag{1.7}$$

If I dissolve the $a^{\mu}(G)$ discrete shift :

$$
\begin{aligned}
+ap &= \left[\tilde{G}\overline{Q},\overline{\tilde{G}}Q\right]_{-} = \left[\sum_{i=1}^{\infty}\varepsilon_i\overline{Q},\sum_{j=1}^{\infty}\overline{\varepsilon}_jQ\right]_{-} \\
&= \sum_{i,j=1}^{\infty}\left[\varepsilon_i\overline{Q},\overline{\varepsilon}_jQ\right]_{-} \neq \sum_{i=1}^{\infty}\left[\varepsilon_i\overline{Q},\overline{\varepsilon}_iQ\right]_{-} = p\sum_{i=1}^{\infty}\delta_{a_i}
\end{aligned}
\qquad (1.8)
$$

The (measureable) anticommutator of these sums is different from the sum of anticommutators. The decomposition is not measurable. The measurable +a quantum shift should be *discrete leap.*

In general the discrete symmetries break the invariance laws, and continuous symmetries keep the invariance laws, so the quantum leap breaks the invariance laws.

### 1.1 The $\Phi\Phi^{+}$ superfield propagator

The supersymmetrical quantum leap appears in the super propagator and vertex, because I can put this phase in the supersymmetric Lagrange density and it stays invariant.
J. Wess obtained the bosonical superfield propagator from the free Wess Lagrangian ref.2:

The **scalar** superfield propagator and the components of superfield:

$$
\begin{aligned}
&\langle 0|T\{\Phi(y,\Theta)\Phi^{+}(y',\Theta')\}|0\rangle = \langle 0|T\{[A(y)+\sqrt{2}\Theta\psi(y)+\Theta\Theta F(y)]\times \\
&[A^{*}(y')+\sqrt{2}\overline{\Theta'\psi}(y')+\overline{\Theta'\Theta'}F^{*}(y')]\}|0\rangle = \\
&\overline{\Theta'\Theta'}\Theta\Theta\langle 0|T\{FF^{*}\}|0\rangle + \langle 0|T\{AA^{*}\}|0\rangle + 2\overline{\Theta'^{\beta}}\Theta^{\alpha}\langle 0|T\{\psi_{\alpha}\overline{\psi}_{\beta}\}|0\rangle
\end{aligned}
\qquad (1.9)
$$

Substitute the components, and use $y = x + i\Theta\sigma\overline{\Theta}$ and $y^{+} = x - i\Theta\sigma\overline{\Theta}$, we see that this propagator has the following x, x' dependence:

$$
\begin{aligned}
&\langle 0|T\{\Phi(x_1,\Theta_1,\overline{\Theta_1})\Phi^{+}(x_2,\Theta_2,\overline{\Theta_2})\}|0\rangle = \\
&= -i\exp[i(\Theta_1\sigma^{m}\overline{\Theta}_1 + \Theta_2\sigma^{m}\overline{\Theta}_2 - 2\Theta_1\sigma^{m}\overline{\Theta_2})\partial_m]\Delta_F(x_2 - x_1)
\end{aligned}
\qquad (1.10)
$$

Write the Goldstino field $\Theta_2 = \tilde{G}$ and initial state $\Theta_1 = 0$ in eq. 1.10!
We get similar discrete space-time translation, like to the definition of SUSY.
The non continuous 4D particle path is available. It likes to the teleportation: neither vertex, nor information exchange during the teleportation. The entangled pair of teleportation is the bosonical superfield. In stop decay entangled pair is the top quark and the Goldstino; the Goldstino is extra dimensional dark matter.

The propagator of self adjoint **vector** superfield ($V=V^+$) has no quantum leap phase, ref. 5:

$$\langle 0|T\{V(x_1,\Theta_1,\overline{\Theta_1})V(x_2,\Theta_2,\overline{\Theta_2})\}|0\rangle = -\frac{1}{8\pi^2}\Delta_F(x_2-x_1)\delta(\Theta_1-\Theta_2) \tag{1.11}$$

From the super Yang-Mills the gluon can't disappear, the Grassman space of gluino can't change, because the Dirac delta Grassman dependence.
The Gluino (svector) can not emit Goldstino, so it decays first into squark. The mass difference between stop and gluino is small, so Gluino decay is slow: long life of Gluino.

## 1.2 **The Goldstino-fermion-boson vertex**

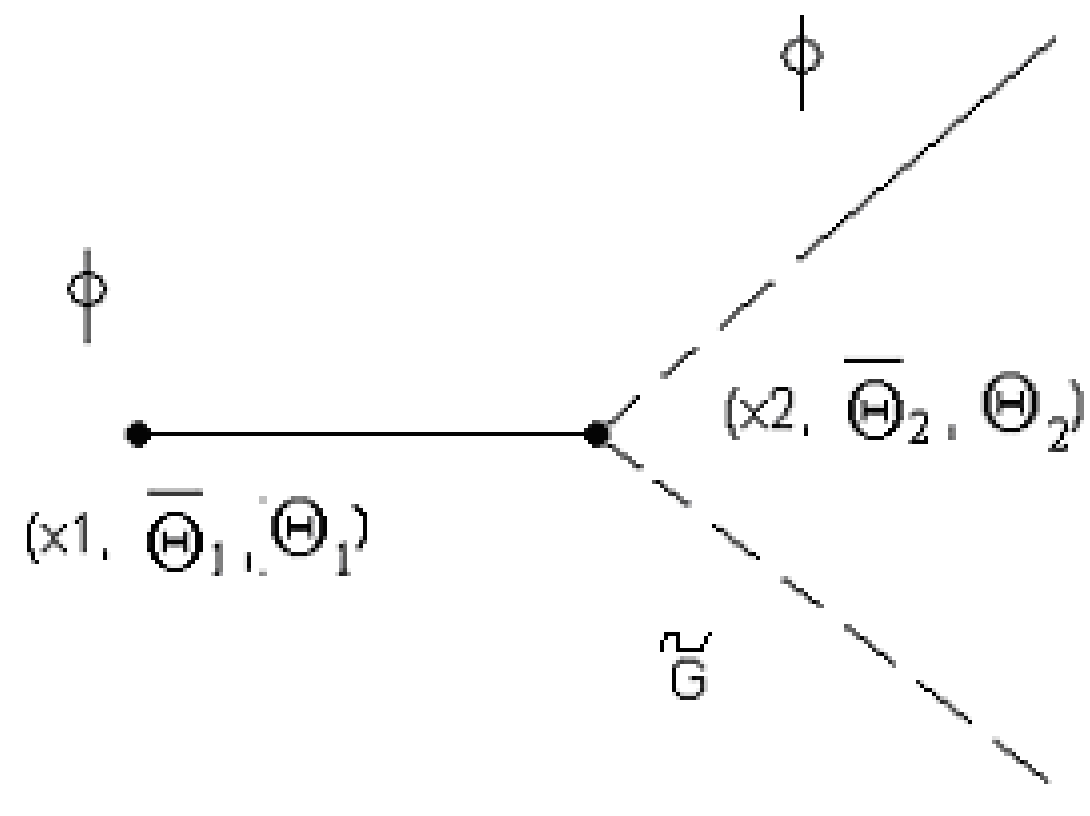

Fig. 1.
$e^{ipa}$ space shift (dashed line) in the squark decay
Local color breaking (quark disappearance) in the squark decay. The dashed line can't interact for a discrete time.

The first description of stop decay is the derivate coupling of Goldstino and super current (like to $p^+ \to n^0 + \pi^+$):

$$L^{\text{int}}_{derived} = \frac{1}{F}\partial_\mu \tilde{G}^a J^\mu_a + h.c. \tag{1.12}$$

where F is the decay constant, J is the supercurrent. This Lagrangian is equivalent to the following nonderivate Goldstino interaction Lagrange, where the Goldberg-Treiman relation fixes the goldstino-boson-fermion coupling to the mass difference of boson and fermion in ref. 4. The following non-derived interacting Lagrangian describe the squark decay:

$$L^{sQCD}_{non-derived} = \frac{m^2_{\tilde{q}} - m^2_q}{F}\tilde{G}q\tilde{q}* - \frac{g_{strong}m_{\tilde{g}}}{\sqrt{2}F}\tilde{G}\tilde{g}^a\tilde{q}^*_i T^a_{ij}\tilde{q}_j \tag{1.13}$$

where $\tilde{q}$ is the squark, q represents the quark, $\tilde{g}$ is a gluino, $\tilde{G}$ is the Goldstino, $g_{Strong}$ is gauge coupling and $T_{ij}^{a}$ is the generator. The new $\tilde{q} \rightarrow q + \tilde{G}$ vertex:

$$\Gamma_{q,\tilde{q},G} = \frac{m_{\tilde{q}}^{2} - m_{q}^{2}}{F} \gamma_{\mu} \delta_{i,j} \tag{1.14}$$

So the vertex:

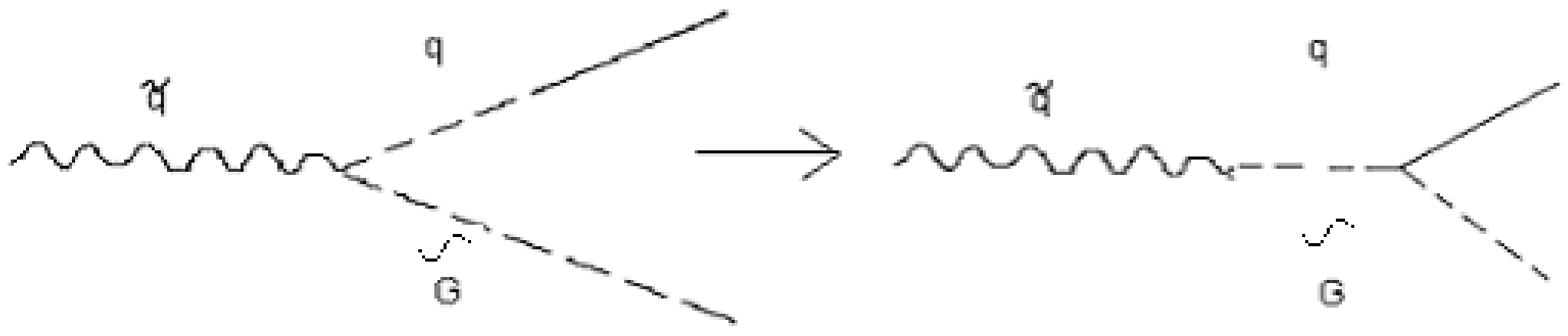

fig.2

The $a^{\mu}$ quantum shift of boson and fermion is equal, they reappear in the same space-time point, where q can absorb the emitted $\tilde{G}$ if it has zero virtual impulse.
Let's see the one loop virtual correction to the superfield propagator:

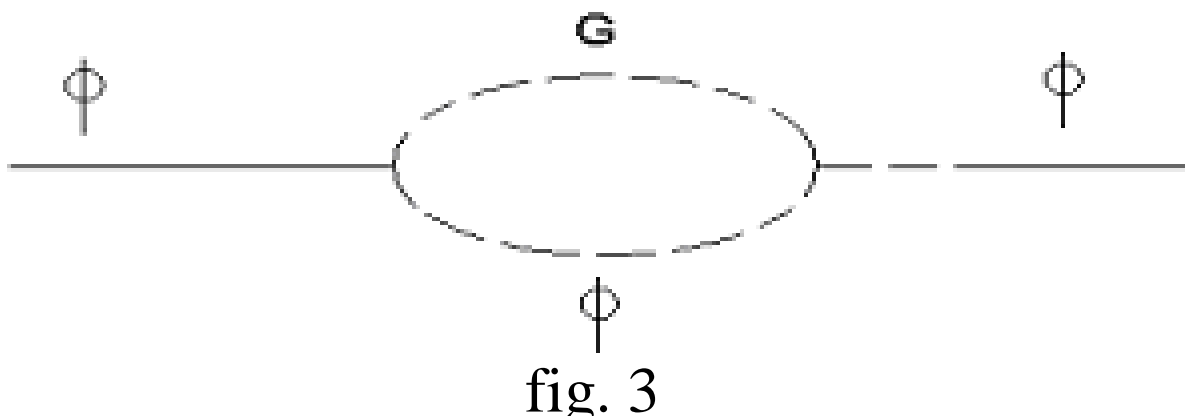

fig. 3

The dashed line Φ is quantum shifted, so the fermion and scalar components have the same quantum leap

The **scatter matrix** of the quark teleportation:

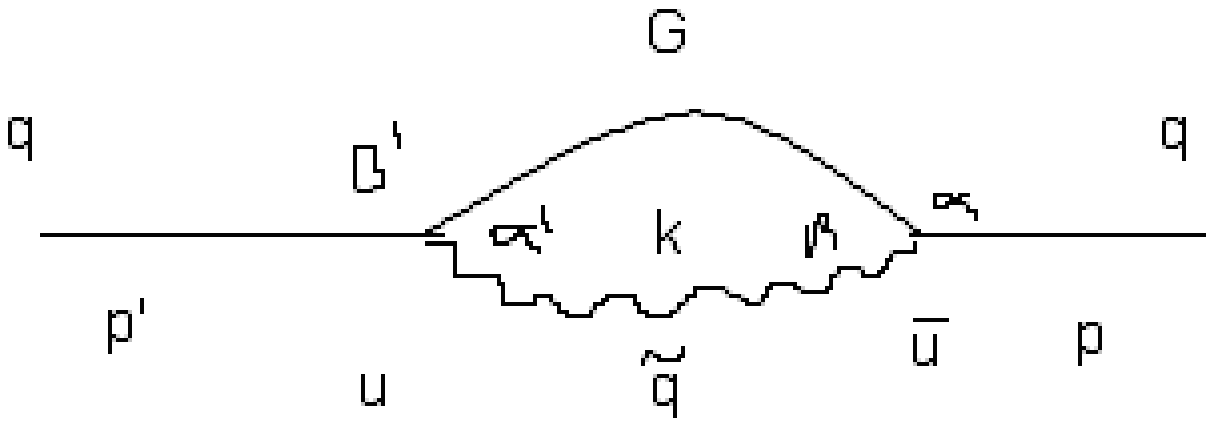

fig. 4.

$$S = \frac{\bar{u}(p')}{(2\pi)^{3/2}} \frac{u(p)}{(2\pi)^{3/2}} \int d^4k/(2\pi)^4 \frac{g_{\mu\nu}(-i)}{k^2 - m_{sq}^2} e^{ika} \frac{1}{k_{Gm}\gamma^m - m_G} e^{ik_G a}$$

$$\{(-i)(2\pi)^4 \frac{m_{sq}^2 - m_q^2}{F} \gamma_\nu \delta^4(p'-k-k_G)(-i)(2\pi)^4 \frac{m_{sq}^2 - m_q^2}{F} \gamma_\mu \delta^4(p-k-k_G)\} \qquad (1.15)$$

$$= (\frac{m_{sq}^2 - m_q^2}{F})^2 (2\pi i)\bar{u}(p')[\frac{1}{(p-k_G)^2 - m_{sq}^2} e^{ipa} \frac{1}{k_{Gm}\gamma^m - m_G}]u(p)\delta^4(p-p')$$

The outgoing field is quantum shifted: $u(p)e^{ipa}$. The component field of superfield propagator contain space-time shift. The coupling is strong, because the mass difference is high. *If the particle energy exceeds the Goldstone fermion mass, then the quantum leap will work.*

$$e^{ipa} = e^{i\tilde{G}\bar{Q} + i\bar{\tilde{G}}Q} = 1 + \bar{\tilde{G}} 2i\sigma^\mu \tilde{G}\partial_\mu + \ldots \approx 1 + 2j^\mu_{\tilde{G}}\partial_\mu \qquad (1.16)$$

The power series contains Goldstino spin current in spin change decay.

*Note*: single Q is hermitian operator only on the Goldstino Hilbert space.

$$\delta_\varepsilon \tilde{G} = \varepsilon/\kappa + i\kappa(\varepsilon\sigma^m \bar{\tilde{G}} - \tilde{G}\sigma^m \bar{\varepsilon})\partial_m \tilde{G} \qquad (1.17)$$

The Goldstino is shifted by $\varepsilon/\kappa$ and quantum shifted by $a^m = i\kappa(\varepsilon\sigma^m \bar{\tilde{G}} - \tilde{G}\sigma^m \bar{\varepsilon})$. Any change in Goldstino wave cause a quantum leap, so it is not observable.

## 2. The QCD response to the local color breaking of SUSY

If a lone color charge disappears today, the whole universe would turn into a not white QGP again, like $10^{10}$ years ago in the Big Bang. I examined the infinite color potential of the color charge breaking. The color antiscreening forbids the escape of quarks and gluons from the hadrons, and forbids the existence of free quarks.
The equal definitions of **quark confinement**:
1. the not color singlet state has got infinite energy,
2. at infinite distance the singlet quark- antiquark potential becomes infinite,
3. the gluon spectrum has mass-gap on low energy,
4. the color charge rises with the distance.
5. Singlet states: $\left|Barion\right\rangle = \frac{1}{\sqrt{6}} \varepsilon^{i,j,k} q^i q^j q^k$ $\qquad \left|Meson\right\rangle = \frac{1}{\sqrt{2}} \delta^{i,j} q^i \bar{q}^j$
6. The renormalisation group confinement exists only if:
$Z^{-1}_3 = 0$ ; $N_{flavor} \le N_{color} + 1$ ; the Lie group is not broken and non-Abelian. And bound state exists with anticommutating FP ghosts.

The static quark potential is linear on large distance:

$$V\ (r) = -k(\alpha_S)\ r \tag{2.0}$$

The "k" constant depends on the $\alpha_S(\Delta q)$ strong coupling constant, and $\alpha_S(\Delta q)$ depends on the impulse difference (running coupling constant):

$$k(\alpha_S) = \frac{3}{5}\alpha_S(q) M^2_{gap}(q) \tag{2.1}$$

$M_{gap}$ is the gluon mass gap in the confined hadron, it is in relationship with the effective range of gluon. The linear (spring) potential bounds the quarks and orders the gluons in narrow flux tubes.

If we create (or disappear) an extra quark color, the non singlet potential energy in eq.2.0 would be around infinity, it is the first definition of quark confinement. The "spring" potential connects **all** quarks.

The infinite bound energy *approach* is the integral of the "k" number.

$$E = \sum_{j}^{Sources} \sum_{i}^{quarks} \int_0^{r_{ij}} dr(-k(\alpha_S(q^2_{ij}))) \rightarrow -\infty \tag{2.2}$$

Where N is the number of the quarks of the universe and $r_i$ is their place. In QGP k=0 means the free plasma state. So we get again a very dense, hot and charged QGP universe. k=0 is a compulsion, because k>0 rises the energy. In the free plasma the *range of gluons* is $\infty$, and $M^{gap}_{gluon} \rightarrow 0$. This extra color charge polarizes all other hadrons and accelerate them until $k(\alpha_S(q^2)) \rightarrow 0$.

The pressure in the V=-kr color potential collapse the distant hadrons into a not white QGP. The color breaking would be cosmic catastrophe.

For example: an anti red quark disappears then the color charge begins to rise with the distance. The lone red quark color seems infinite red from large distance. This red attracts the blue and green quarks of any nucleon and repulses the red quark of the nucleons. But the quark color charge is random in the hadrons, so this potential attracts with 2/3-1/3 force the hadrons. The white hadrons can't neutralize this extra red color. The red quark accelerate the hadrons until the momentum difference becomes $q$, where $k(\alpha_S(q))$=0.

The re-appearance of the missing anti red charge(s) dissolve the infinite bound energy and the gluon range confined to some fermi; $M_{gap}$>0; k>0. The globally white QGP gas can expand and cool.

## 3 The re-Big Bang:

1. The quantum leap breaks the charge and energy for a discrete time.
2. Free gluons expand, connect all quarks and accelerate them. The not white state has got infinite energy. The total color becomes a random parameter, because more and more squark disappear. In the accelerated world the particles become massless in EW restoration, the vev. of Higgs boson vanish. The universe contracts into a color QGP sphere.
3. The baryon genesis (exponential B number grow) and matter - antimatter breaking needs long time (>sec) and therefore color broken state. Because the sphaleron and GUT baryon number breaking rates are slow. The $t_{2x}$ time of Baryon number doubling on high temperature is about a second. So in the standard Big Bang theory, where nothing was before the Big Bang, nothing would be in 3 sec.

$$B = B_0 2^{\frac{t}{t_{2x}}} \tag{3.1}$$

$$\begin{aligned} 1/t_{sph} &= \kappa \alpha_{Weak}^4 T \\ 1/t_{GUT} &= \kappa \alpha_{GUT}^2 T \end{aligned} \tag{3.2}$$

4. The plasma is able to explode, if the color restore to neutral white. The restoration is solvable in strongly curved space-time, because the quantum shift is not hermitian in SUGRA. The squarks can't disappear more.

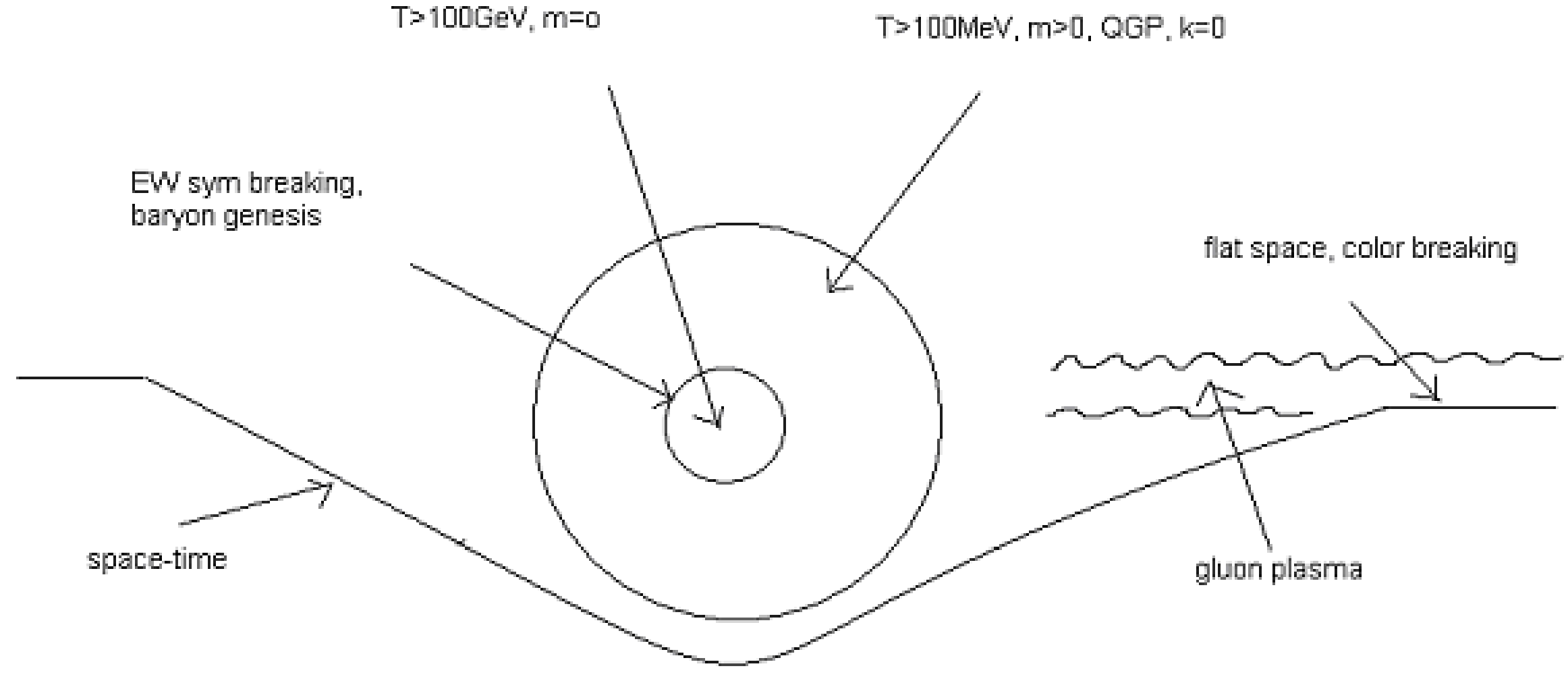

Fig 5.

5. The inflation begins if all free charges are in the near of the plasma, and are in curved space-time.
6. The explosion is not irreversible only if: at first the plasma cool below the confinement and all squarks decay, and after these, the space is flat. Then Universe can continue the inflation.
7. 6 need equithermal particles in the expansion.
   The temperature: $T \sim 1/R^3$ and the metric $g \sim m/R$; Where the curvature became negligible $g \sim 0$ there the radius of flat space is large, and T could be smaller if m is bigger:

$$T \xrightarrow{space \to flat} T_{confinement} \qquad (3.3)$$

   In other words, the huge mass Universe is earlier confined and later flat.

## 4. Consistence with the observations

- If a vertex contains Goldstino, then there are quantum shifted outgoing legs.
- If the lightest SUSY particle is the Goldstino it will *vanish forever,* because with self interactions we can add infinite amount of time shifts $G(t + \sum a_0)$. Goldstino is the lightest supersymmetrical particle and is not observable, like to **dark matter.**
- In the nature don't may exist supersymmetric QGP on flat space-time.
- In the near of the black holes the space time is strongly curved $\left[x_1, x_2\right] \neq 0$.

In SUGRA the double SUSY transformation of vector superfield changes to:

$$(\delta_\eta \delta_\varepsilon - \delta_\varepsilon \delta_\eta) V^A = V^D \varepsilon^C \eta^B R^A_{BCD} - \varepsilon^C \eta^B T^D_{BC} D_D V^A \qquad (4.1)$$

These equations are more complicated because new gravitational fields appear. The **curved SUSY shift isn't hermitian operator.** In curved space the super particle can not vanish, so the black holes eat the dark matter.
The covariant derivate of bosons and fermions in SUGRA contains the gravitino product with superpartner:

$$\Phi_i = A_i + \sqrt{2}\Theta\chi_i + \Theta\Theta F_i \qquad (4.2)$$

$$D_a A^*_i = e_a{}^m \partial_m A^*_i - \frac{\sqrt{2}i}{2} \overline{\psi}_{a\dot{\kappa}} \overline{\chi}_i^{\dot{\kappa}} \qquad (4.3)$$

$$D_a \overline{\chi}_{i\dot{\alpha}} = e_a{}^m D'_m \overline{\chi}_{i\dot{\alpha}} + \frac{\sqrt{2}i}{2} \overline{\psi}_\alpha{}^\beta \sigma^b_{\beta\dot{\alpha}} D_b A^*_i - \frac{\sqrt{2}}{2} \overline{\psi}_{a\dot{\alpha}} F^*_i \qquad (4.4)$$

The covariant derivate and so the double SUSY transformation is not measurable in SUGRA.
- The confined space of black holes confines the quarks, gluons can't leave the event horizon.

- Only the gravitational interaction of Goldstino doesn't give time shifts. Dark matter gravitates continuously. If **previous Big Bangs** disperse the dark matter in a large volume, the Hubble expansion could be faster: more dark matter is out of the Universe, then inside the observable Universe. Only in the curved space of the black holes is the Goldstino real.
- On the centre of a Galaxy is a huge black hole, the spiral arm rotation speed is proportional with the distance from centre, and this means anomalous gravitation on galactic distances. So astronomy introduces the dark matter, and dark energy because anomalous increasing Hubble constant is observed in Astronomy.

- The *cosmic ray* energy of protons and light ions is $10^{12}$-$10^{20}$ eV.
The beam is a light single ion; the target is a solid rock or gas/plasma. So quantum leap must be short << fm.

- Fermions and bosons have equal mass $m=0$ during the disappeared translation, this is the same mass multiplet of unbroken SUSY. Fermions and bosons have equal occupation of states because $T=0$. (But before and after the disappearance these particles have high temperature.)
- I explain the flat Universe with the collapsed Universe (like supernova explosion). The spherical QGP collapse and flow out by a flat plane ($J_{in}=J_{out\ flow}$).
- I explain the homogenous microwave background with the long random color Universum, a short Big Bang would be turbulent, and not homogenous.
-The slow QGP collapse and inflation has enough time to reach the thermal equilibrium state, so the matter density and $\gamma$ background is homogeneous.
- The literature of SUSY find the dark matter = $\chi^0$ , which has no interaction with matter, but why is the cross section = 0 ??? I explain it with quantum leap of Goldstino particle.

## 5. Calculus

a) The main question is the length of *quantum leap*; it is almost real vector.

The Grassman odd $\Theta_1$ has a dimension $\frac{1}{\sqrt{Mass}}$ .

$$a^{\mu} = \overline{\Theta}_1 2i\sigma^{\mu}\Theta_1 \qquad (1.4)$$

$$\sigma^0 = \begin{bmatrix} 1 & 0 \\ 0 & 1 \end{bmatrix}, \quad \sigma^1 = \begin{bmatrix} 0 & -1 \\ -1 & 0 \end{bmatrix}, \quad \sigma^2 = \begin{bmatrix} 0 & i \\ -i & 0 \end{bmatrix}, \quad \sigma^3 = \begin{bmatrix} -1 & 0 \\ 0 & 1 \end{bmatrix}$$

The Pauli matrices are real, except $\sigma^2$. So only the $i\sigma^2$ matrix is imagine.

If I would like to create a real space shift with dimension [meter] from $\overline{\Theta}\Theta$ [$\frac{1}{M}$], I must take the De Broglie wavelength of a 160 GeV Goldstino: $\lambda = \frac{hc}{pc}$ ~ 10^-18 m.

De Broglie wavelength can be dangerous if the Goldstino mass zero is, all the masses were zero before SUSY/EW breaking, so in Big Bang was Goldstino dangerous light.

b) *Estimate the LSP (Goldstino) mass* from dark matter mass in universe.
The mass ratio after the Big Bang: 0 % dark energy, 63% dark matter, 10% neutrinos, 15% photons, 12% atoms. I assumed, that the hadron production has 100 pb and the sparticle production has 1 pb total cross section.

Then the ratio $\frac{12\%}{63\%} = \frac{100 * 0{,}3 GeV}{M GeV}$ $M_{goldstinc}$~157 GeV $\pm 100\%$

The mass of lightest squark is about: $M_{stop}$>157+175 GeV = 332 GeV

Before SUSY breaking the goldstino mass was 0, the squark and quark mass is equal, and the sq decay is dangerous, because the long De Broglie wavelength. But now the stop is heavy and the decay is safe.

If the Goldstino mass is much less: ~ $m_e$, then it carriers the most energy from the decay, and so a longer De Broglie wavelength, which makes the decay subcritical. (Impulse conservation)

c) *Chain reaction*. An extra color quark attract strong the another quarks, and collide with all. The quarks fly aimed, the fluence is very high. The fluence is calculated from a color accelerated quarks of lead target ( $N_{quark} = 10^{29} \frac{1}{m^3}$ ), where one color breaking was.

The reaction:

$$R = \sigma \psi \tag{5.1}$$

$$\psi = \frac{1}{4\pi r^2} \frac{dN(t)}{dt} = \frac{1}{4\pi\lambda^2} \frac{4\pi\lambda^2 c dt}{dt} * 10^{29} \approx 3 * 10^{37} \frac{1}{m^2 s} \tag{5.2}$$

$\sigma_{stop-pair}$=1 pb

$R \approx 10^{19} \frac{1}{s}$ is supercritical, if energy > $m_{stop-pair}$

d) Accelerating *color potential* is 0,8 GeV / fm so 1 TeV need 10^-12 m ! Ref (6)
The quantum leap< 10^-12 m: **the chain reaction can not start**.

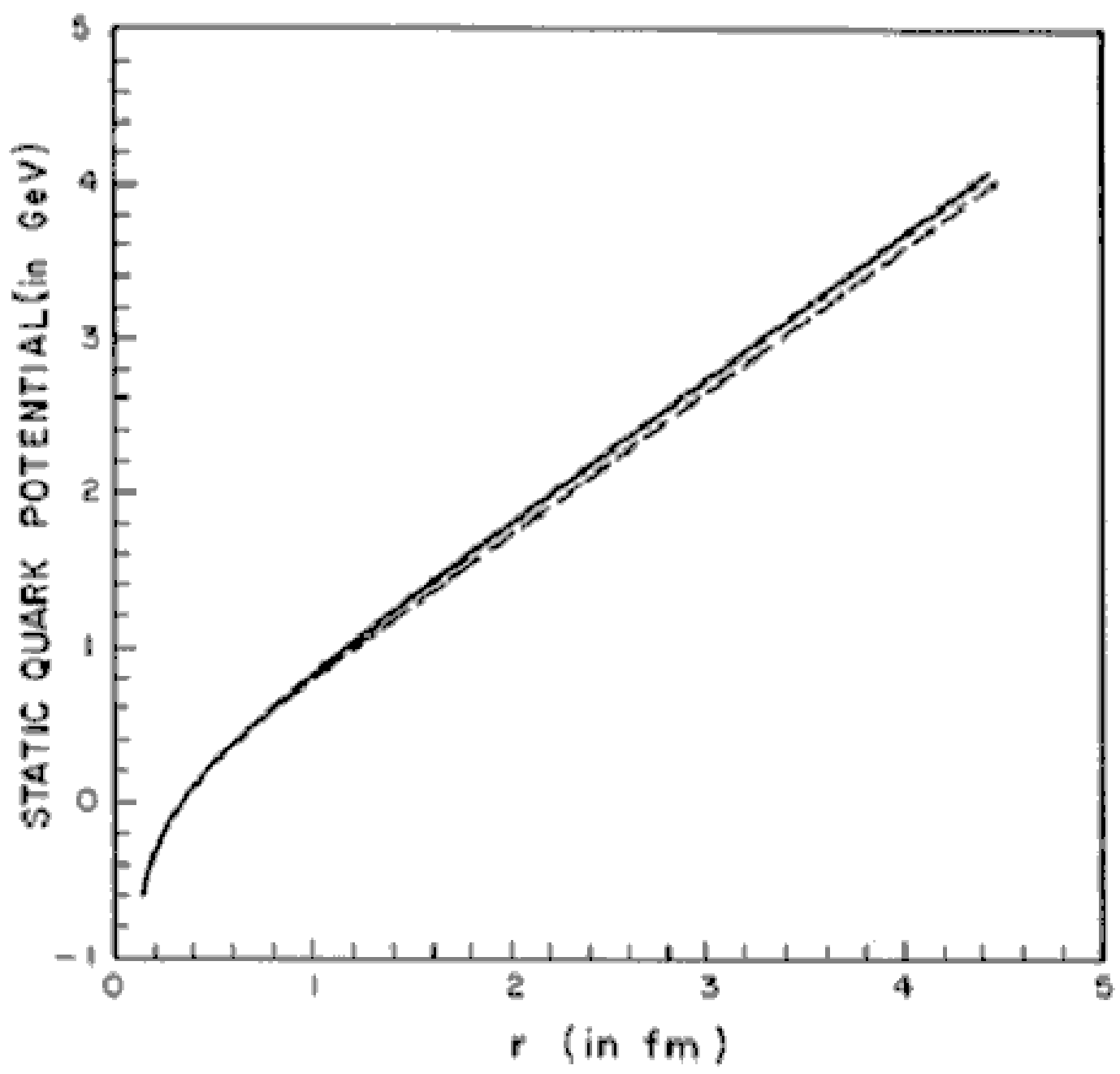

Static quark-antiquark potential as a function of source separation.

e) If I assume, that before the big bang was only one meson, than which meson potential has the energy of universe?

   The gravitation and kinetic energy of universe is equal, together about zero. The mass (10^55 kg) has ~10^91 eV energy. It's equal with 10^67 meter biquark potential. It would be bigger as our space time.
   The radius of universe is about 4*10^26 m. So before the big bang was more color breaking, not only one.

## 6. Classical Grassman mechanics

The simple conclusion of the classical Grassman mechanics from Ref. 7. (Gerhard Soff: Supersymmetrie)
The real Lagrange of the free Grassman particle:

$$L = \frac{i}{2}\Theta(t)\dot{\Theta}(t) \tag{6.1}$$

The classical trajectory of the free Grassman particle is constant state or coordinate:

$$\dot{\Theta}(t) = 0 \tag{6.2}$$

The Lagrange of the Grassman particle with Source:

$$L = \frac{i}{2}\Theta(t)\dot{\Theta}(t) - i\Theta(t)\eta \tag{6.3}$$

The classical trajectory with source:

$$\dot{\Theta}(t) = \eta \tag{6.4}$$

The Grassman space need a source to change. The source strength prescribes the Grassman shift and the teleportation length in space-time.
The canonical momentum is the coordinate:

$$\pi = -\frac{i}{2}\Theta(t) \tag{6.5}$$

The quantum Grassman mechanics has a Grassmann Position operator (N) and a canonical conjugated momentum (iP), which antikommutator:

$$[N, P]_{-} = 1 \tag{6.6}$$

The open question is: Are the colliders producing supersymmetry charge?
Because some charges $[\overline{Q}, Q]_{-}$ in QGP can make long color breaking and a chain reaction.

Ref.: